\documentclass[12pt]{article}
\usepackage[pdftex,usenames,dvipsnames]{color}
\usepackage{graphics}
\usepackage{amsfonts}

\newcommand{\nc}{\newcommand}
\nc{\blue}{\textcolor{Blue}}
\nc{\navy}{\textcolor{NavyBlue}}
\nc{\cyan}{\textcolor{Cyan}}
\nc{\violet}{\textcolor{Violet}}
\nc{\purple}{\textcolor{Purple}}
\nc{\red}{\textcolor{Red}}
\nc{\sienna}{\textcolor{RawSienna}}
\nc{\brown}{\textcolor{Brown}}
\nc{\orange}{\textcolor{Orange}}
\nc{\yellow}{\textcolor{Yellow}}
\nc{\green}{\textcolor{Green}}
\nc{\olive}{\textcolor{OliveGreen}}

\newcommand{\nit}{\noindent}

\newcommand{\np}{\newpage}
\newcommand{\dsp}{\displaystyle}
\newcommand{\vs}[1]{\vspace{#1 ex}}
\newcommand{\hs}[1]{\hspace{#1 em}}
\newcommand{\bflr}{\begin{flushright}}
\newcommand{\eflr}{\end{flushright}}
\newcommand{\bc}{\begin{center}}
\newcommand{\ec}{\end{center}}
\newcommand{\ben}{\begin{enumerate}}
\newcommand{\een}{\end{enumerate}}

\newcommand{\be}{\begin{equation}}
\newcommand{\ee}{\end{equation}}
\newcommand{\ba}{\begin{array}}
\newcommand{\ea}{\end{array}}
\newcommand{\ct}{\cite}
\newcommand{\bit}{\bibitem}
\newcommand{\dd}[2]{\frac{\partial{#1}}{\partial{#2}}}
\newcommand{\ag}{\alpha}

\newcommand{\gam}{\gamma}
\newcommand{\del}{\delta}
\newcommand{\eps}{\epsilon}
\newcommand{\ve}{\varepsilon}

\newcommand{\kg}{\kappa}
\newcommand{\lb}{\lambda}
\newcommand{\sg}{\sigma}

\newcommand{\vf}{\varphi}
\newcommand{\og}{\omega}
\newcommand{\Gam}{\Gamma}
\newcommand{\Del}{\Delta} 

\newcommand{\Fg}{\Phi}

\newcommand{\lh}{\left(}
\newcommand{\rh}{\right)}

\newcommand{\rd}{\right.}

\newcommand{\bfa}{{\bf a}}

\newcommand{\bfn}{{\bf n}}

\newcommand{\bfv}{{\bf v}}
\newcommand{\bfA}{{\bf A}}
\newcommand{\bfB}{{\bf B}}
\newcommand{\bfE}{{\bf E}}

\newcommand{\bfX}{{\bf X}}

\newcommand{\cE}{{\cal E}}

\newcommand{\der}{\partial}

\begin{document}
\pagestyle{empty}

\begin{flushright}
NIKHEF/2008-009
\end{flushright}
\vs{7}

\bc
{\Large {\bf The gravitational field of a light wave}} \\
\vs{5}

{\large J.W.\ van Holten}\\
\vs{3}

{\large NIKHEF, Amsterdam NL}
\vs{4}

\today
\ec
\vs{10}

\nit
{\small
{\bf Abstract} \\
According to the classical Einstein-Maxwell theory of gravity and 
electromagnetism, a light-wave traveling in empty space-time is accompanied 
by a gravitational field of the {\em pp}-type. Therefore point masses are 
scattered by a light wave, even if they carry no electric or magnetic charge, 
or dipole moment. In this paper I present the explicit form of the metric and 
curvature for both circularly and linearly polarized light, and discuss the 
geodesic motion of test masses. This is followed by a discussion of classical
scattering of point particles by the gravitational field associated with a 
circularly polarized electromagnetic block wave. A generalization to a 
quantum theory of particles in the background of these classical wave fields 
is presented in terms of the covariant Klein-Gordon equation. I derive the
energy spectrum of quantum particles in the specific case of the circularly 
polarized block wave. Finally, a few general remarks on the extension to a 
quantum light wave are presented. 
}

\np 
\pagestyle{plain}
\pagenumbering{arabic}

\section{Introduction \label{s.1}}

General Relativity provides an excellent account of all known gravitational
phenomena, such as planetary orbits, gravitational lensing and the dynamics
of binary neutron stars. It also predicts the existence of black holes and
gravitational waves, which has motivated intense efforts of physicists and 
astronomers to observe their presence and properties. 

To observe gravitational phenomena in the laboratory is much more difficult,
as a result of the weakness of gravity as compared to other fields of force,
in particular electromagnetism, but also strong nuclear interactions and
even the weak interactions of quarks and leptons mediated by the massive 
vector bosons $(Z, W^{\pm})$. Gravitational experiments in the laboratory
are mostly confined to measurements of Newton's constant and tests of the
equivalence principle, although the gravitational redshift has also
been established in terrestrial experiments and gravitational time-dilation
is nowadays of practical importance for the accuracy of satellite-based 
global position measurements. 

The gravitational fields involved in these tests of General Relativity are 
almost always provided by large masses, such as that of the earth, the sun 
and other stars or massive compact objects. However, as Einstein's theory
tells us that all forms of energy are a source of gravitational fields,
it is of some interest to study the gravitational field associated with 
wave-phenomena, such as electromagnetic waves, as well as certain related
purely gravitational waves. Solutions of General Relativity and of 
Einstein-Maxwell theory describing these physical situations are known to 
exist \ct{brinkmann}-\ct{jwvh}. In this paper I study their properties and 
analyze the dynamics of massive and massless test particles in the presence 
of these fields, paying in particular attention to the gravitational effects. 

\section{Waves in Einstein-Maxwell theory \label{s.2}}

In the absence of massive particles and electric charges or currents, the 
combined theory of gravitational and electromagnetic fields is specified 
by the Einstein-Maxwell equations
\be 
\ba{l}
R_{\mu\nu} - \frac{1}{2}\, g_{\mu\nu} R = - 8\pi G\, T_{\mu\nu}, \\
 \\
D^{\mu} F_{\mu\nu} = 0, 
\ea
\label{2.1}
\ee
where the energy-momentum tensor of the electromagnetic field is
\be
T_{\mu\nu} = F_{\mu\lb} F_{\nu}^{\;\;\lb} - \frac{1}{4}\, g_{\mu\nu} F^2,
\label{2.2}
\ee
$F^2$ being the trace of the first term, and where we have chosen units 
such that $c = \ve_0 = 1$. In edition to the equations (\ref{2.1}) there are
Bianchi identities, implying for the Maxwell tensor $F_{\mu\nu}$ that it 
can be written in terms of a vector potential $A_{\mu}$:
\be
F_{\mu\nu} = \der_{\mu} A_{\nu} - \der_{\nu} A_{\mu},
\label{2.3}
\ee
and for the Riemann tensor that it can be expressed in terms of the metric 
via a torsion-free symmetric connection. 

The Maxwell equations can be solved in empty Minkowski space-time in 
terms of plane waves. Such plane waves are characterized by a constant 
wave vector and electric and magnetic field strengths which are orthogonal 
to the wave vector and to each other, as in fig.\ 1.  

\bc
\scalebox{0.5}{\hs{12}\includegraphics{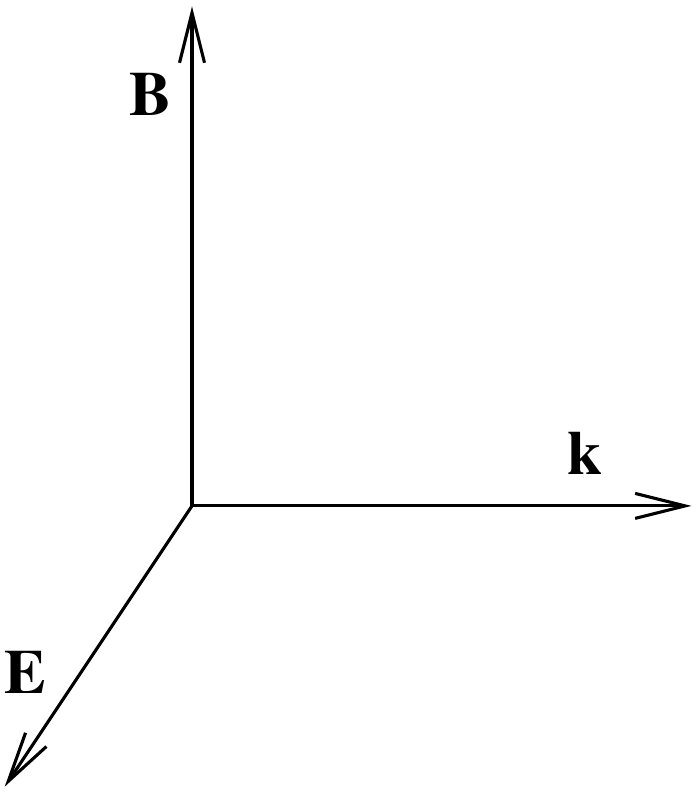}} 
\vs{1}

{\footnotesize{Fig.\ 1: Electric and magnetic field strength of a 
harmonic plane wave.}}
\ec

\nit 
The properties of plane-wave solutions must be preserved in the full
Einstein-Maxwell theory, in particular there should exist covariantly
constant light-like wave vector fields
\be 
k^2 = g_{\mu\nu} k^{\mu} k^{\nu} = 0, \hs{2} D_{\mu} k_{\nu} = 0,
\label{2.7}
\ee
and electromagnetic vector potentials $A_{\mu}$ which describe waves
traveling at the velocity of light, and which are locally orthogonal 
to the wave vector field $k^{\mu}$. Such solutions exist \ct{papapetrou}
and can be cast in the following form. In $4$-dimensional space-time we 
introduce light-cone co-ordinates
\be 
u = t - z, \hs{1} v = t + z. 
\label{2.4}
\ee
Then the vector potential, written as a 1-form, for waves traveling 
in the positive $z$-direction is 
\be
A = A_i(u) dx^i,
\label{2.8}
\ee
where $x^i$ are the co-ordinates in the 2-dimensional transverse 
plane, and the components of the vector potential can be expanded in 
plane waves:
\be 
A_i(u) = \int \frac{dk}{2\pi}\, \lh a_i(k) \sin ku + b_i(k) \cos ku \rh.
\label{2.9}
\ee
The corresponding field strength $F = 2 dA$ has components
\be
F =  2 F_{ui}\, du \wedge dx^i 
 = 2 A^{\prime}_i(u)\, du \wedge dx^i ,
\label{2.10}
\ee
from which it follows that the electric and magnetic fields are 
transverse, taking the values 
\be
E_i = - \ve_{ij} B_j = F_{ui}(u) = A^{\prime}_i(u).
\label{2.11}
\ee
In flat space-time the energy-momentum tensor of this electromagnetic
field reads
\be
T_{\mu\nu}\, dx^{\mu} dx^{\nu} = T_{uu}\, du^2, \hs{2}
T_{uu} = F_{ui} F_{u}^{\;\;i} 
 = \frac{1}{2} \lh \bfE^2 + \bfB^2 \rh.
\label{2.12}
\ee
In view of eq.\ (\ref{2.1}) the Ricci tensor is required to have the 
same form; this holds for space-times with a metric of the {\em pp}-wave 
type:
\be
g_{\mu\nu} dx^{\mu} dx^{\nu} = - du dv - \Fg(u,x^i) du^2 + dx^{i\,2}.
\label{2.13}
\ee
The explicit components of the connection and the Riemann tensor are 
presented in the appendix. Here it suffices to notice, that the Ricci
tensor is of the required type indeed:
\be
R_{\mu\nu} dx^{\mu} dx^{\nu} = R_{uu} du^2, \hs{2}
R_{uu} = -\frac{1}{2}\, \der_i^2\, \Fg.
\label{2.15}
\ee
The metric (\ref{2.13}) admits a constant light-like Killing vector
\be 
K = k^{\mu} \der_{\mu} = 2 k \der_v,
\label{2.14}
\ee
signifying the translation invariance of all fields in the $v$-direction.

Substitution of the Ricci tensor (\ref{2.15}) and the electro-magnetic
energy-momentum tensor (\ref{2.12}) in the Einstein equations leads to
a single non-trivial field equation
\be 
\der_i^2\Fg = 8 \pi G \lh \bfE^2 + \bfB^2 \rh. 
\label{2.16}
\ee
Moreover, the explicit form of electro-magentic field strength tensor
(\ref{2.10}) and the connection coefficients (\ref{a.1}) guarantees that
the Maxwell equations 
\be
D^{\mu} F_{\mu\nu}= 0,
\label{2.17}
\ee
reduce to the same equations in Minkoswki space, and therefore hold 
for the vector potentials (\ref{2.8}), (\ref{2.9}).

Similar {\em pp}-wave solutions in the presence of non-gravitational
fields can be constructed for scalar and Dirac fields \ct{skmh,jwvh}, 
Yang-Mills fields \ct{fuster-vh} and higher-rank antisymmetric tensors 
as in 10-D supergravity \ct{blau-et-al,maldacena-maoz,fuster}. 
Dimensional reduction of {\em pp}-waves has been used to construct
explicit solutions of lower-dimensional non-relativistic field theories
\ct{duval-horv-palla}.

\section{{\em PP}-wave solutions \label{s.3}}

As the electric and magnetic fields $\bfE(u)$, $\bfB(u)$ in 
eq.\ (\ref{2.16}) depend only on the light-cone variable $u$, the
equation can be integrated to give the result
\be 
\Fg = 2 \pi G\, (x^2 + y^2) \lh \bfE^2 + \bfB^2 \rh + \Fg_0,
\label{3.1}
\ee
where $\Fg_0$ is a solution of the homogeneous equation
\be
\der_i^2 \Fg_0 = 0. 
\label{3.2}
\ee
Trivial solutions of the homogeneous equation are represented by 
the linear expressions
\be
\Fg_0(u,x^i) = \Fg_{flat}(u,x^i) = \ag(u) + \ag_i(u) x^i.
\label{3.3}
\ee
Since the Riemann tensor is composed only of second derivatives 
$\Fg_{,ij}$, these linear solutions by themselves describe a flat 
space-time with $R_{\mu\nu\kg\lb} = 0$. The metric only looks 
non-trivial because it describes Minkowski space as seen from an 
accelerated co-ordinate system.  

In general the number of non-trivial quadratic and higher solutions 
depends on the dimensionality of the space-time. In 4-$D$ space-time 
there are two linearly independent quadratic solutions:
\be
\Fg_0(u,x^i) = \Fg_{gw}(u,x^i) 
 = \kg_+(u) \lh x^2 - y^2 \rh + 2 \kg_{\times}(u) x y,
\label{3.4}
\ee
where $\kg_{+,\times}(u)$ are the amplitudes of the two polarization
modes. The Riemann tensor now has non-vanishing components
\be
R_{uxux} = - R_{uyuy} = - \kg_+(u), \hs{2}
R_{uxuy} = R_{uyux} = - \kg_{\times}(u). 
\label{3.5}
\ee
If these amplitudes are well-behaved, the solutions represent 
non-singular gravitational-wave space-times. We can also infer that 
these modes have spin-2 behaviour; indeed, under a rotation around 
the $z$-axis represented by the co-ordinate transformation
\be 
\lh \ba{c} 
    x^{\prime} \\ y^{\prime} \ea \rh 
    = \lh \ba{cc} 
     \cos \vf & - \sin \vf \\
     \sin \vf & \cos \vf \ea \rh \lh \ba{c} x \\ y \ea \rh,  
\label{3.6}
\ee
$\Fg_{gw}$ is invariant if we simultaneously perform a rotation 
between the amplitudes
\be 
\lh \ba{c} \kg_+^{\prime} \\ \kg_{\times}^{\prime} \ea \rh = 
 \lh \ba{cc} \cos 2 \vf & - \sin 2 \vf \\
            \sin 2 \vf & \cos 2 \vf \ea \rh
 \lh \ba{c} \kg_+ \\ \kg_{\times} \ea \rh. 
\label{3.7}
\ee 
Note that the form of the special solution (\ref{3.1}) implies,
that with this rule the full metric is invariant under rotations 
in the transverse plane.  

For all integers $n > 2$ there also exist non-trivial solutions 
$\Fg_0$ constructed from linear combinations of $n$-th order monomials 
in $x^i$. They form a spin-$n$ representation of the transverse rotation 
group $SO(2)$. However, all such solutions give rise to singular curvature 
components at spatial infinity. Thus the spin-2 solutions (\ref{3.4}) 
seem to be the only globally {\em bona fide} free gravitational wave 
solutions of this kind, and we restrict the space of the {\em pp}-wave 
solutions of the Einstein-Maxwell equations to the line elements defined 
by the solution (\ref{3.1}) with $\Fg_0 = \Fg_{gw}$ as in (\ref{3.4}). 

\section{Geodesics \label{s.4}}

Returning to the em-wave space-times (\ref{3.1}), the geodesics 
can be determined from the connection coefficients (\ref{a.1}), 
given by the components of the gradient of $\Fg$. As a result, 
the relevant equations reduce to those of a particle moving in 
a potential, as we now show \ct{duval-horv-palla,jwvh}.

Consider time-like geodesics $X^{\mu}(\tau)$, parametrized by 
the proper time $\tau$:
\be
d\tau^2 = dU dV + \Fg(U,X^i) dU^2 - dX^{i\,2}.
\label{4.1}
\ee
This choice of parameter immediately establishes the square of the 
4-velocity as a constant of motion:
\be
g_{\mu\nu}(X)\, \dot{X}^{\mu} \dot{X}^{\nu} = 
 - \dot{U} \dot{V} - \Fg(U,X^i)\, \dot{U}^2 + \dot{X}^{i\,2} = - 1,
\label{4.2}
\ee
where the overdot denotes a proper-time derivative. This equation
gives a first integral of motion for the light-cone co-ordinate 
$V(\tau)$ in terms of solutions for the other three co-ordinates. 
These have to be obtained from the geodesic equation
\be 
\ddot{X}^{\mu} + \Gam_{\lb\nu}^{\;\;\;\mu} \dot{X}^{\lb} \dot{X}^{\nu} = 0.
\label{4.2.1}
\ee
The existence of the Killing vector (\ref{2.14}) implies a 
simple equation for the other light-cone co-ordinate $U(\tau)$:
\be
\ddot{U} = 0 \hs{1} \Rightarrow \hs{1} 
\dot{U}(\tau) = \gam = \mbox{constant},
\label{4.3}
\ee
as there are no connection coefficients with contravariant index 
$\mu = u$. It follows, that $U$ can be used to parametrize geodesics, 
instead of $\tau$. 

Now eq.\ (\ref{4.2}) implies for the laboratory time co-ordinate 
$T = X^0$:
\be
\frac{dT}{d\tau} = \sqrt{\frac{1 - \gam^2 \Fg}{1 - \bfv^2}},
\label{4.4}
\ee
where $\bfv = d \bfX/dT$ is the velocity in laboratory 
co-ordinates\footnote{Note, that our notation here is not covariant: 
\[
\bfv^2 = \sum_{a=1}^3 v_a^2 \neq \sum_{a,b = 1}^3 g_{ab} v^a v^b.
\]}
Now, as
\be 
\frac{dU}{dT} = 1 - v_z, 
\label{4.5}
\ee
it follows from eqs.\ (\ref{4.2}) and (\ref{4.4}) that $h$ defined by
\be 
h \equiv \frac{1 - \bfv^2}{(1 - v_z)^2} + \Fg = \frac{1}{\gam^2},
\label{4.6}
\ee
is a constant of motion, the gravitational equivalent of the total
particle energy. In particular, for a particle initially at rest in 
a locally flat space-time one finds $h = \gam = 1$. 

For light-like geodesics one can follow a similar procedure by 
introducing an affine parameter $\lb$ such that the geodesic 
equation (\ref{4.2.1}) holds upon interpreting the overdot as a 
derivative w.r.t.\ $\lb$, whilst the line element and the left-hand 
side of eq.\ (\ref{4.2}) are taken to vanish. It then follows, that
\be
h = \frac{1 - \bfv^2}{(1 - v_z)^2} + \Fg = 0.
\label{4.7}
\ee
Observe in particular that, as $\bfv^2$ is not a covariant expression, 
in general $\bfv^2 \neq 1$, even for light. 

Next we turn to the transverse part of the motion. Considering either
time-like or light-like geodesics, the geodesic equation (\ref{4.2.1}) 
for the transverse co-ordinates $X^i$ takes the form
\be
\ddot{X}^i = - \frac{\gam^2}{2}\, \Fg_{,i} \hs{1} \Leftrightarrow \hs{1}
\frac{\der^2 X^i}{\der U^2} + \frac{1}{2}\, \dd{\Fg}{X^i} = 0.
\label{4.8}
\ee
For a quadratic potential it follows that
\be 
\Fg = \kg_{ij}(U)\, X^i X^j \hs{1} \Rightarrow \hs{1}
\dd{^2 X^i}{U^2} + \kg_{ij}(U)\, X^j = 0. 
\label{4.9}
\ee
This is the equation for a parametric oscillator with real or
imaginary frequencies, depending on the signs of the components
$\kg_{ij}(U)$. 

A special case is that of negative constant curvature; after a 
diagonalization of the coefficients $\kg_{ij}$ this situation is 
characterized by (temporarily suspending the summation convention):
\be 
R_{uiuj} = - \kg_{ij} \equiv - \mu_i^2\, \del_{ij} 
\hs{1} \Rightarrow \hs{1} X^i(U) = X_0^i\, \cos \mu_i (U - U_0).
\label{4.10}
\ee
The remarkable aspect of this result, is that the magnitude of the
curvature (the gravitational field strength) determines the
{\em frequency} of the geodesic motion, rather than its amplitude. 
This represents a gravitational analogue of the Josephson
effect, where a constant voltage generates an oscillating current.

\section{The gravitational field of a light wave \label{s.5}}
 
In section \ref{s.2} we discussed general wave solutions of the 
Einstein-Maxwell theory. We now consider the special case of
monochromatic waves. As our first example we take the light wave 
to be circularly polarized; the corresponding vector potential 
can be written as
\be 
\bfA = a \lh \cos ku, \sin ku, 0 \rh \hs{1} \Rightarrow \hs{1}
\left\{ \ba{l} \bfE = ka \lh - \sin ku, \cos ku, 0 \rh, \\
       \\
       \bfB = ka \lh \cos ku, \sin ku, 0 \rh. \ea \rd
\label{5.1}
\ee
As the electric and magnetic fields are $90^{\circ}$
out of phase the energy density is constant, and
\be
\Fg_{circ} = 2 \pi G \lh \bfE^2 + \bfB^2 \rh \lh x^2 + y^2 \rh 
 = 4\pi G\, k^2 a^2 \lh x^2 + y^2 \rh.
\label{5.2}
\ee
Thus the potential is of the quadratic type (\ref{4.9}), (\ref{4.10}), 
with
\be 
\mu_x^2 = \mu_y^2 = \mu^2 \equiv 4 \pi G\, k^2 a^2. 
\label{5.3}
\ee
Numerically, we find in SI units
\be 
\mu = 1.3 \times 10^{-9}\, \frac{E}{E_c}\;  \mbox{m$^{-1}$}, \hs{2}
E_c = \frac{m_e^2}{e} = 1.3 \times 10^{18}\; \mbox{Vm$^{-1}$}. 
\label{5.4}
\ee
Here $E_c$ is the critical field for electron-positron pair production. 
For the limiting value $E = E_c$ we find an angular frequency of 
$\og = \mu c = 0.4$ rad/s.

As a second example we take linearly polarized light, for which
\be 
\bfA = a \lh \cos ku, 0, 0 \rh \hs{1} \Rightarrow \hs{1} 
\left\{ \ba{l} \bfE = ka \lh - \sin ku, 0, 0 \rh, \\
       \\
       \bfB = ka \lh 0, \sin ku, 0 \rh. \ea \rd
\label{5.5}
\ee
Then the potential takes the form
\be 
\Fg_{lin} = 4\pi G\, k^2 a^2\, \sin^2 ku \lh x^2 + y^2 \rh.
\label{5.6}
\ee
Introducing the time variable $s = kU$, the transverse equations of motion
becomes
\be
\frac{d^2 X^i}{ds^2} + \nu^2 \lh 1 - \cos 2s \rh X^i = 0,
\label{5.7}
\ee
with
\be 
\nu^2 = 2\pi G\, a^2.
\label{5.8}
\ee
This is a Mathieu equation, with Bloch-type periodic solutions 
\be
\ba{l}
X^i(s) = u(s) \cos qs + v(s) \sin qs, \\
 \\
u(s+\pi) = u(s), \hs{1} v(s+\pi) = v(s).
\ea
\label{5.9}
\ee
For very large $\nu^2$, with electromagnetic field intensities of
the order of the Planck scale, the wave numbers $q$ become complex 
and the solutions exhibit parametric resonance. 

\section{Scattering by the gravitational wave field \label{s.6}}

As a light-wave (\ref{2.8}), (\ref{2.9}) is accompanied by a 
gravitational field of the {\em pp}-type, and as gravity is a 
universal force, even electrically and magnetically neutral 
particles are scattered by a light-wave. Although this 
gravitational force acts on charged particles as well, their 
dynamics is generally dominated by the Lorentz force, depending 
on the ratio of charge to mass. In this section I discuss the 
scattering of neutral classical point particles by wave-like
gravitational fields. 

The general solution to this scattering problem is provided by 
the geodesics discussed in sect.\ \ref{s.4}. However, a more 
generic scattering situation is specified by taking both the 
initial and final states of a particle to be states of inertial 
motion in flat Minkowski space-time, the state of motion being 
changed at intermediate times by the passage of a wave-like 
gravitational field of finite extent. 

For simplicity, let us consider a circularly polarized electromagnetic
block wave, accompanied by a gravitational block wave as sketched
in Fig.\ 2:

\bc
\scalebox{0.75}{\includegraphics{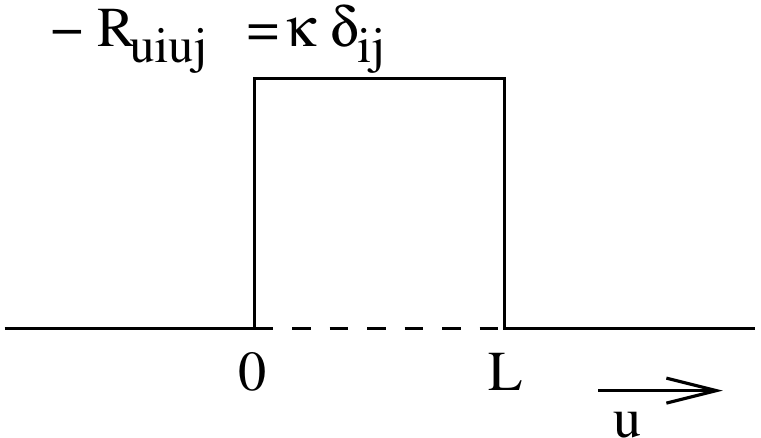}} \\
\vs{1}

{\footnotesize Fig.\ 2: Curvature block wave.}
\ec

\be
\Fg = \kg(u) \lh x^2 + y^2 \rh, \hs{1}
\kg(u) = \left\{ \ba{ll}
                0, & u < 0; \\
		\mu^2, & 0 \leq u \leq L; \\
		0, & u > L. \ea \rd
\label{6.1}
\ee
Now in the asymptotic regions the time-like geodesics are straight lines:
\be
\ba{ll}
X^i(U) = X^i_0 + p^i U,  & U < 0; \\
 & \\
X^i(U) = \bar{X}^i_0 + \bar{p}^i U, & U > L.
\ea
\label{6.2}
\ee
The interpolating solution of oscillating type (\ref{4.10}) must match
these asymptotic solutions at $U = 0$ and $U = L$, such that both 
$X^i(U)$ and $X^{i\,\prime}(U)$ are continuous. By matching at $U = 0$
one gets:
\be 
X^i(U) = X^i_0\, \cos \mu U + \frac{p^i}{\mu}\, \sin \mu U, 
\label{6.3}
\ee
Then matching with the Minkowski-solution for $U > L$ the
following linear relations between the transverse position and 
velocity set-offs are found:
\be 
\lh \ba{c} \bar{X}^i_0 \\ \\ \bar{p}^i \ea \rh = 
 \lh \ba{cc} \cos \mu L + \mu L \sin \mu L & 
    L \lh \frac{\sin \mu L}{\mu L} - \cos \mu L \rh \\ 
  & \\
 - \mu \sin \mu L & \cos \mu L \ea \rh \lh \ba{c} X^i_0 \\ \\ p^i \ea \rh,
\label{6.4}
\ee
In particular, if a particle is initially at rest: $\bfv = 0$,
hence $p_x = p_y = 0$ and $h = \gam = 1$, then the final velocity 
$\bar{\bfv}$ is:
\be 
\ba{l}
\dsp{ \bar{v}_x = \frac{d\bar{X}}{d\bar{T}} 
 = - \frac{\mu X_0 \sin \mu L}{1 + \ag}, \hs{1}
\bar{v}_y = \frac{d\bar{Y}}{d\bar{T}}
 = - \frac{\mu Y_0 \sin \mu L}{1 + \ag}, }\\
 \\
\dsp{ \bar{v}_z = \frac{d\bar{Z}}{d\bar{T}} = \frac{\ag}{1 + \ag}, }
\ea
\label{6.5}
\ee
with
\be 
\ag = \frac{\mu^2}{2} \lh X_0^2 + Y_0^2 \rh \sin^2 \mu L. 
\label{6.6}
\ee
Observe, that the particle will be at rest again in the final state
if $\mu L = n \pi$, with $n$ integer, whereas the transverse velocity
after scattering is maximal for $\mu L = (n + 1/2) \pi$. 
By the results (\ref{6.5}) the scattering angle is given by
\be 
\tan \psi = \frac{\sqrt{\bar{v}_x^2 + \bar{v}_y^2}}{\bar{v}_z}
 = \sqrt{\frac{2}{\ag}},
\label{6.7}
\ee
and therefore $\tan \psi$ is large for small $\ag$, i.e.\ $\mu L \ll 1$.
In contrast, for large $\ag$ the transverse velocity vanishes: 
$\bar{v}_i \rightarrow 0$, whilst the velocity in the $z$-direction 
approaches the speed of light: $\bar{v}_z \rightarrow 1$, hence
$\psi \rightarrow 0$. 

A similar analysis can be done for light-like geodesics, for which 
$h = 0$. If a massless particle, like a photon, initially travels in 
the $-z$ direction with velocity $\bfv = (0, 0, -1)$, and therefore 
$p_x = p_y = 0$, one finds for the final velocity vector $\bar{\bfv}$:
\be 
\ba{l}
\dsp{ \bar{v}_x = \frac{d\bar{X}}{d\bar{T}} 
 = - \frac{2\mu X_0 \sin \mu L}{2\ag + 1}, \hs{1}
\bar{v}_y = \frac{d\bar{Y}}{d\bar{T}}
 = - \frac{2\mu Y_0 \sin \mu L}{2\ag + 1}, }\\
 \\
\dsp{ \bar{v}_z = \frac{d\bar{Z}}{d\bar{T}} = \frac{2\ag - 1}{2\ag + 1}, }
\ea
\label{6.8}
\ee
The corresponding scattering angle for massless particles is
\be 
\tan \psi = \left| \frac{\sqrt{\bar{v}_x^2 + \bar{v}_y^2}}{\bar{v}_z} \right|
 = \frac{2 \sqrt{2 \ag}}{1 - 2 \ag}. 
\label{6/9}
\ee
Observe, that for $\ag = 1/2$ the velocity of massless particles is purely
transverse, whilst for $\ag > 1/2$ the sign of $\bar{v}_z$ reverses, and 
its value approaches $+ 1$ at large transverse distances. 

\section{Quantum fields in a {\em pp}-wave background \label{s.7}}

Like classical particles, also quantum fields are affected by the presence 
of the gravitational field of a light wave, even in the absence of direct
electromagnetic interactions. One can observe this in the behaviour of 
a scalar field in a {\em pp}-wave background, described by the metric 
(\ref{2.13}). The d'Alembert operator then takes the form
\be 
\Box_{pp} = \frac{1}{\sqrt{-g}} \der_{\mu} \sqrt{-g} g^{\mu\nu} \der_{\nu}
 = - 4 \der_u \der_v + 4 \Fg(u;x^i)\, \der_v^2 + \Del_{trans},
\label{7.1}
\ee
where $\Del_{trans} = \sum_i \der_i^2$ is the Laplace operator in the 
transverse plane. The Klein-Gordon equation
\be 
\lh - \Box_{pp} + m^2 \rh \Psi = 0,  
\label{7.2}
\ee
can be transformed by a Fourier transformation
\be 
\Psi(u,v; x^i) = \frac{1}{2\pi}\, \int ds dq\, \psi(s,q; x^i)\, 
e^{-i(su + qv)}.
\label{7.3}
\ee
The equation for $\psi(s,q;x^i)$ then becomes
\be 
\lh - \Del_{trans} + 4 q^2 \Fg(-i \der_s; x^i) - 4 qs + m^2 \rh \psi = 0.
\label{7.4}
\ee
In the special case of quadratic $\Fg$ with constant curvature, as in eqs.\
(\ref{4.9}) and (\ref{4.10}):
\be
\Fg = \mu^2_x x^2 + \mu_y y^2,
\label{7.5}
\ee
the equation can be solved in closed form, in terms of hermite polynomials
$H_n(x)$. More generally, introducing the ladder operators 
\be 
\bfa_i = \frac{1}{2\sqrt{|q|\mu_i}} \lh \der_i + 2|q|\mu_i\, x_i \rh, \hs{1}
\bfa_i^{\dagger} = \frac{1}{2\sqrt{|q|\mu_i}} \lh - \der_i + 2|q|\mu_i\, 
x_i\rh,
\label{7.6}
\ee
with commutation relations
\be 
\left[ \bfa_i, \bfa_j^{\dagger} \right] = \del_{ij},
\label{7.7}
\ee
the Klein-Gordon equation (\ref{7.4}) becomes
\be
\left[ \sum_{i = (x,y)} 4 |q| \mu_i 
 \lh \bfa_i^{\dagger} \bfa_i + \frac{1}{2} \rh - 4qs + m^2 \right] \psi = 0.
\label{7.8}
\ee
Now write 
\be 
E = s + q, \hs{1} p = s - q \hs{1} \Rightarrow \hs{1}
s u + q v = Et - pz;
\label{7.9}
\ee
then the integer eigenvalues of the occupation number operator
$\bfn_i = \bfa_i^{\dagger} \bfa_i$ turn the Klein-Gordon equation 
into an equation for the spectrum of energy eigenvalues for
the scalar field:
\be 
\lh E \mp \sg \rh^2 = \lh p \mp \sg \rh^2 +m^2, \hs{1}
\sg(n_i) = \sum_i \mu_i \lh n_i + \frac{1}{2} \rh 
         \geq \frac{1}{2}\, \sum_i \mu_i,
\label{7.10}
\ee
where the sign depends on whether $E > p$ (upper sign), or $E < p$
(lower sign). The levels $\sg$ are quantized, the $n_i$ being 
non-negative integers. Taking into account (\ref{7.8}) the general 
solution for the Klein-Gordon equation can be written in explicit 
form as
\be 
\ba{l}
\dsp{ \Psi(u,v;x^i) = \frac{1}{2\pi}\, \sum_{n_i = 0}^{\infty} 
\int_{-\infty}^{\infty} ds \int_{-\infty}^{\infty} dq\, 
\del(4sq - 4 \sg |q| - m^2)\, \chi_{n_i}(q)\, e^{- i su - i vq} }\\
 \\
\hs{7} \dsp{ \times\, \prod_{j = x,y} \left[ 
\lh \frac{2\mu_j|q|}{\pi} \rh^{1/4}
\frac{H_{n_j}(\xi_j)}{\sqrt{2^{n_j} n_j!}}\, 
e^{-\xi_j^2 /2} \right], }
\ea
\label{7.11}
\ee
where
\be 
\xi_i = \sqrt{2 \mu_i |q|}\, x_i.
\label{7.12.0}
\ee
Performing the integral over $s$ and taking $\Psi$ to be real, this takes 
the form
\be 
\ba{l}
\dsp{ \Psi(u,v;x^i) = \frac{1}{2\pi}\, \sum_{n_i = 0}^{\infty} 
\int_0^{\infty} \frac{dq}{q}\, \lh a_{n_i}(q)\, e^{-iqv - i \lh \frac{m^2}{4q}
+ \sg \rh u} + a^*_{n_i}(q)\, e^{iqv + i \lh \frac{m^2}{4q} + \sg \rh u} \rh }\\
 \\
\hs{7} \dsp{ \times\, \prod_{j = x,y} \left[ 
\lh \frac{2\mu_j|q|}{\pi} \rh^{1/4}
\frac{H_{n_j}(\xi_j)}{\sqrt{2^{n_j} n_j!}}\, 
e^{-\xi_j^2 /2} \right], }
\ea
\label{7.12}
\ee
with the reality condition resulting in
\be
a_{n_i}(q) = \frac{1}{4}\, \chi_{n_i}(q), \hs{1} 
a_{n_i}^*(q) = - \frac{1}{4}\, \chi_{n_i}(-q), \hs{2} q > 0.
\label{7.13}
\ee
The Fock space of the quantum scalar field is then generated by 
taking the Fourier coefficients to be operators with commutation
relation
\be 
\left[ a_{n_i}(q), a^*_{m_i}(k) \right] = \pi |q|\,
 \del_{n_x,m_x} \del_{n_y,m_y}\, \del (q - k).
\label{7.14}
\ee
Equivalently, the space-time fields themselves then obey the equal 
light-cone time commutation relation 
\be 
\left[ \Psi(u,v,x^i), \Psi(u,v^{\prime},x^{\prime\,i}) \right]
 = \frac{i}{4}\, \eps(v - v^{\prime})\, \del^2\lh x^i - x^{\prime\, i} \rh,
\label{7.15}
\ee
with $\eps(x)$ the Heavyside step function
\be 
\eps(x) = \left\{ \ba{cl} +1, & x > 0; \\
                        -1, & x < 0. \ea \rd
\label{7.16}
\ee
From expression (\ref{7.12}) we read off that the lowest single-particle 
energy is $E = m + \sg(0)$, for $p = \sg(0)$.

\section{Discussion \label{s.8}}

In this paper I have presented the properties of the gravitational field 
associated with a light wave. The effects of this gravitational field are 
extremely small, but qualitatively and conceptually very interesting. 

So far I have described the light-wave as a classical Maxwell field;
however, ultimately one would like to consider a quantum description 
of light and its associated gravitational effects. To see what kind of 
issues are at stake, let me momentarily restore SI units, and summarize 
the equations for the electromagnetic energy density and flux, and the 
corresponding space-time curvature:
\be 
\Fg = c \cE = \frac{\ve_0 c}{2} \lh E^2 + B^2 \rh 
 = - \frac{c^5}{8\pi G}\, R_{uu}.
\label{8.1}
\ee
Now according to the quantum theory as first developed by Planck and 
Einstein, we can also think of the wave in terms of photons of energy
$\hbar \og$. Then the flux is expressed in terms of photons per unit 
of time and area as
\be 
\Fg = \hbar \og\, \frac{dN}{dt dA}.
\label{8.2}
\ee
Equating the two expressions above, we get the relationship
\be 
\frac{1}{\og}\, \frac{dN}{dt} = -\frac{R_{uu}}{k^2}\, \frac{dA}{l_{Pl}^2},
\hs{2} l_{Pl}^2 = \frac{8 \pi G \hbar}{c^3}.
\label{8.3}
\ee
Both sides of this equation represent dimensionless numbers, with the
right-hand side actually a product of two dimensionless quantities:
the ratio of Ricci curvature $R_{uu}$ and wavenumber squared $k^2 = 
4\pi^2/\lb^2$, and the area $dA$ in Planck units.  At the microscopic 
level at least some, and possibly all, of these quantities have to 
exhibit quantized behaviour.

\appendix

\section{Appendix: \\ connection and curvature tensor}

Starting fromt the metric (\ref{2.13}), the connection can be computed.
The only non-zero connection coefficients are 
\be
\Gam_{uu}^{\;\;\;v} = \Fg_{,u}, \hs{1} 
\Gam_{iu}^{\;\;\;v} = 2 \Gam_{uu}^{\;\;\;i} = \Fg_{,i},
\label{a.1}
\ee
and the Riemann tensor reduces to the components
\be 
R_{iuj}^{\;\;\;\;\;v} = -2 R_{uiu}^{\;\;\;\;\;j} = \Fg_{,ij}.
\label{a.2}
\ee
These equations are equivalent to a single equation for the completely
covariant components of the Riemann tensor:
\be 
R_{uiuj} = - \frac{1}{2}\, \Fg_{,ij}.
\label{a.3}
\ee
From eq.\ (\ref{a.2}) one directly finds the Ricci tensor (\ref{2.15}).
The Riemann scalar obviously vanishes: $R = 0$. 

\np

\end{document}